\documentclass[%
superscriptaddress,
eprint,
twocolumn,
amsmath,
amssymb,
prx,
]{revtex4-2}

\usepackage[english]{babel}
\usepackage{graphicx}% Include figure files
\usepackage{dcolumn}% Align table columns on decimal point
\usepackage{bm}% bold math
\usepackage{dsfont}
\usepackage[colorlinks]{hyperref}% add hypertext capabilities
\hypersetup{%
    plainpages=true,
    breaklinks=true,
    hypertexnames=false,
    pageanchor=true,
    colorlinks=true,
    linkcolor={blue},
    citecolor={blue},
    urlcolor={blue},
    anchorcolor={black}
}
\usepackage{hhline}
\usepackage{braket, siunitx}
\usepackage{ifthen}
\newboolean{showcomments}
\setboolean{showcomments}{true}

\usepackage{cleveref}
\usepackage{xcolor}
\usepackage{bbding} % for checkmarks
\usepackage{comment}
\usepackage{xspace}
\usepackage{hyperref}
\usepackage{soul}

\crefname{equation}{Eq.}{Eqs.}
\Crefname{equation}{Equation}{Equations}
\crefname{figure}{Fig.}{Figs.} 
\Crefname{figure}{Figure}{Figures}
\crefname{section}{Sect.}{Sects.}
\Crefname{section}{Section}{Sections}
\crefname{table}{Table}{Tables}
\crefname{appsec}{}{Appendices} 

\newcommand{\mynote}[3]{%
  \ifthenelse{\boolean{showcomments}}{%
   \fbox{\bfseries\sffamily\scriptsize#1}%
   {\small$\blacktriangleright$\textsf{\emph{\color{#3}{#2}}}$\blacktriangleleft$}}%
  {%
   \@bsphack
   \@esphack
  }%
}

\newcommand{\panel}[1]{({#1})}

\newcommand*{\figref}[2]{%
  \hyperref[{#1}]{%
    ~\ref*{#1}%
    \ifx\\#2\\%
    \else
      \panel{#2}%
    \fi
  }%
}

\newcommand{\hc}{\mathrm{H.c.}}

\newcommand{\qpdag}[1]{\hat{\gamma}^\dagger_{#1}}
\newcommand{\qpo}[1]{\hat{\gamma}_{#1}}
\newcommand{\ovphi}{\hat{\varphi}}

\newcommand{\qp}{\mathrm{QP}}

\newcommand{\hqpphi}{\hat{H}_{t}}

\newcommand{\xqp}{x_\mathrm{QP}}

\newcommand{\nlb}{\nolinebreak}

\newcommand{\sbar}{\bar s}
\newcommand{\ve}{\varepsilon}

\newcommand{\pat}{\mathrm{PAT}}

\begin{document}
\preprint{APS/123-QED}
\widetext
\title{Numerical Modeling of Quasiparticle-Induced Dissipation in Fluxonium Qubits}

\def\LLaffil{Lincoln Laboratory, Massachusetts Institute of Technology, Lexington, MA 02421, USA}
\def\RLEaffil{Research Laboratory of Electronics, Massachusetts Institute of Technology, Cambridge, MA 02139, USA}
\def\Physaffil{Department of Physics, Massachusetts Institute of Technology, Cambridge, MA 02139, USA}
\def\EECSaffil{Department of Electrical Engineering and Computer Science, Massachusetts Institute of Technology, Cambridge, MA 02139, USA}

\author{Kate Azar}\email{kate.azar@ll.mit.edu}\affiliation{\LLaffil}\affiliation{\RLEaffil}\affiliation{\EECSaffil}
\author{Max Hays}\affiliation{\RLEaffil}
\author{Kyle Serniak}\email{kyle.serniak@ll.mit.edu}\affiliation{\LLaffil}\affiliation{\RLEaffil}

\date{\today}
\begin{abstract}
Nonequilibrium quasiparticles (QPs) generated by stray infrared and ionizing radiation can limit the performance of superconducting quantum processors and present challenges for quantum error correction schemes.
Models of QP-induced energy relaxation commonly assume that the characteristic energy of the QPs and the qubit transition energy are both small relative to the superconducting gap.
Under these assumptions, certain qubits such as the fluxonium would exhibit protection against QP-induced dissipation at specific bias points.
Here, we show that this is not necessarily the case, numerically analyzing the predicted rate of QP-induced dissipation in fluxonium qubits for different QP energy distributions and for QPs created via photon-assisted tunneling processes.
We find that accounting for small numerical factors, existing theoretical models predict sensitivity to QP-induced errors at bias points previously thought to be protected.
We find that inclusion of asymmetry in the superconducting gap energy across the junction can reintroduce suppression of QP-induced relaxation, as expected.
Additionally, for QPs created by photon-assisted tunneling, we predict that $T_1$ protection will only occur for a specific energy of pair-breaking radiation.
This understanding of fluxonium sensitivity to QP-induced dissipation informs the development of fluxonium-based processors and future QP-mitigation strategies.
\end{abstract}

\maketitle

%%%%%%%%%%%%%%%%%%%%%%%%%%%%%%
\section{Introduction}
Fermionic excitations out of the BCS superconducting ground state, referred to as Bogoliubov quasiparticles (QPs), can induce decoherence in superconducting qubits.
QPs can exchange energy with a qubit, leading to depolarization errors \cite{lutchyn_prb_2005, Martinis_EnergyDecay_2009, catelani_prb_2011, aumentado_quasiparticle_2023,glazman_bogoliubov_2021}, and their presence can suppress the critical current of Josephson junctions (JJs), causing qubit dephasing \cite{catelani_prb_2011, Catelani_2012, Kurilovich_2026}.
These effects have been well-studied in theory and experiments, as the presence of QPs challenges the fundamental assumption that the quantum dynamics of superconducting circuits can be fully described by the macroscopic degrees of freedom of the Cooper-pair condensate \cite{Devoret_1985}, and evidence points to a nonequilibrium population far exceeding expected values at experimental temperatures~\cite{aumentado2004nonequilibrium,shaw_prb_2008,barends2011minimizing,saira_prb_2012,nsanzineza_prl_2014,serniak_hot_2018,serniak_direct_2019,iaia2022phonon}.
More recently, interest in QP effects has intensified due to their role in catastrophic bursts of errors correlated in space and time which pose challenges for quantum error correction \cite{vepsalainen2020impact, McEwen_2022, mcewen2024resisting, harrington2025synchronous}.

Recent efforts to mitigate QP-induced errors in superconducting qubits can be binned into two categories. 
First, one can reduce the population of QPs in sensitive areas of the qubits.
These efforts have seen success from filtering and shielding of QP-generating infrared radiation~\cite{barends2011minimizing,corcoles2011protecting,serniak_thesis_2019}, assessing and mitigating the effects of ionizing radiation in the environment~\cite{vepsalainen2020impact,cardani2021reducing,harrington2025synchronous,loer2024abatement,de2026evaluating}, and ``gap engineering" qubits to trap QPs away from sensitive areas~\cite{riwar2016normal,serniak_thesis_2019}, block QP tunneling at JJs~\cite{aumentado2004nonequilibrium, marchegiani_quasiparticles_2022, mcewen2024resisting,pinckney2026characterization}, and downconvert pair-breaking phonons~\cite{henriques2019phonon,martinis2021saving,iaia2022phonon}.
The second is to use circuits where energy exchange between the qubit degree of freedom and the presumed population of QPs is suppressed at the intended operating points.
In circuits such as the flux qubit \cite{Orlando_1999}, fluxonium qubit \cite{manucharyan_fluxonium_2009}, $\cos 2\varphi$ qubit~\cite{smith2020superconducting}, etc., this intrinsic insensitivity relies on the assumption that QPs exist at energies sufficiently near the superconducting gap edge both before and after the interaction with the qubit~\cite{pop_coherent_2014}.

In this Article, we describe how for most practical qubit circuits the aforementioned assumption is invalid, but that insensitivity to QPs can still be realized via other considerations. 
Using the fluxonium qubit as a case study, we utilize established theoretical descriptions to numerically simulate the energy relaxation rate due to tunneling of a nonequilibrium QP population and from photon-assisted QP tunneling (PAT) processes~\cite{Tien_PAT_1963,houzet_photon-assisted_2019, liu_prl_2024, chowdhury_2026}, taking into account the initial and final energy of QPs above the gap edge.
For simplicity, we first consider two initial energy distributions of QPs: a ``hot" distribution meant to emulate a radiation burst event, and a ``cold" distribution meant to emulate QPs in steady state, which we expect to be well thermalized to the phonon bath (described by an offset chemical potential \cite{Connolly_2024}). 
We predict qubit sensitivity to QPs for both the ``hot" and ``cold" QP distributions when accounting for the energy transfer to and from the QPs. 
When we include an explicit asymmetry of the superconducting gap on either side of the JJ, sensitivity to the ``cold" distribution of QPs is reduced.  
We investigate PAT processes in the fluxonium and find suppressed sensitivity only for a monochromatic drive of a specific pair-breaking frequency.
Combined, these findings reveal a consequential interplay of assumptions that updates previous interpretations of QP-qubit interactions.

\section{QP-Qubit Coupling}
\begin{figure}[t]
    \centering
    \includegraphics[width=\linewidth]{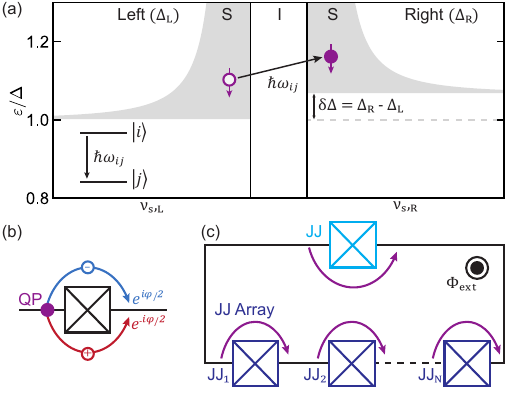}
    \caption{(a) QP density of states in energy around a JJ. A QP is depicted tunneling from the left to the right electrode, acquiring energy $h\omega_{ij}$ from the qubit and inducing the qubit transition from state $\ket{i}$ to state $\ket{j}$. Asymmetry in the superconducting gap across the junction electrodes ($\delta\Delta$) is also represented. (b) Cartoon of a QP residing at the superconducting gap tunneling across a JJ. If the QP were to tunnel from gap edge to gap edge, it would acquire phase $\pm \varphi/2$ with the sign depending on whether it tunnels as an electron (blue) or a hole (red). (c) Diagram of QP tunneling across qubit circuit elements. In a fluxonium circuit, a single JJ (cyan) in parallel with an array of JJs in series (dark blue) create a superconducting loop that can be threaded with a magnetic flux bias $\Phi_{\mathrm{ext}}$. Purple arrows indicate QPs tunneling.}
    \label{fig1}
\end{figure}

QPs couple to the phase degrees of freedom in superconducting circuits.
As such, QPs have the most effect in parts of the circuit where the quantum phase fluctuations are strongest.
In many qubits (including the popular transmon~\cite{koch_2007} and the fluxonium~\cite{manucharyan_fluxonium_2009} considered here) this interaction is strongest at the JJs comprising the circuit. 
We therefore model the interaction between QPs and qubits via tunneling processes across JJs [Fig~\ref{fig1}].
By applying the Bogoliubov-Valatin transformation \cite{Bogoliubov} to the single-electron tunneling Hamiltonian~\cite{Bardeen_tunneling_1961}, we obtain the QP tunneling Hamiltonian described by Catelani et al.~\cite{catelani_prb_2011} and others:
\begin{equation}
\begin{aligned}
    \hqpphi= t\sum_{l,r,s}\bigg[&(u_ru_le^{i\ovphi/2}-v_rv_le^{-i\ovphi/2})\qpdag{rs}\qpo{ls}\\
    &+(u_rv_le^{i\ovphi/2}+v_ru_le^{-i\ovphi/2})\qpdag{rs}\qpdag{l\sbar}\bigg]+\hc
\end{aligned}
\label{eq: factored qp-qubit interaction term}
\end{equation}
%%%%%%%%%%%%%%%%
Here $\hat{\gamma}_{ls}(\hat{\gamma}_{ls}^\dagger)$ is the fermionic QP annihilation (creation) operator corresponding to spin $s$ (opposite spin $\sbar$) in state $l$ ($r$) in the left (right) electrode of the junction.
Assuming no charge imbalance, the coefficients $u_l$ and $v_l$ are given by 
\begin{equation}
    \label{eq:usandvs2}
    u_l^2, v_l^2 = \frac{1}{2}\left(1\pm\frac{\sqrt{\ve_l^2-\Delta_l^2}}{\ve_l}\right),
\end{equation}
where $\Delta_l$ is the superconducting gap energy of the left electrode and $\varepsilon_l$ is the energy of a QP measured from the Fermi energy.

There are two terms in Eq.~\ref{eq: factored qp-qubit interaction term}: one that conserves QP excitation number~($\qpdag{rs}\qpo{ls}$) and another that creates two QP excitations~($\qpdag{rs}\qpdag{l\sbar}$).
This form of the number-conserving term illustrates a common picture of quasiparticle tunneling: the transfer of a QP from left to right by the operators $\qpdag{rs}\qpo{ls}$ is the coherent superposition of two processes --- the QP tunneling ``as an electron" (acquiring phase $\hat\varphi/2$ with probability $u_ru_l$) or ``as a hole"  (acquiring phase $-\hat\varphi/2$ with probability $v_rv_l$) as depicted in ~Fig.~\ref{fig1}~(b). 
For tunneling of QPs at the gap edge (where $u_ru_l=v_rv_l=0.5$), these terms can interfere perfectly destructively when $\varphi=\pm\pi$.

For an isolated qubit with no drives, and for the qubit transition energy between states $\ket{i}$ and $\ket{j}$ $\hbar\omega_{ij}$ less than $2\Delta$, only the QP-number-conserving term in Eq.~\ref{eq: factored qp-qubit interaction term} will contribute to the qubit transition rate, corresponding to the tunneling of preexisting QPs.
However, if there is high frequency radiation with $\hbar\omega_p\gtrsim2\Delta$ in the qubit environment, QP-pair creation can occur at the junction (corresponding to the terms associated with $\qpdag{rs}\qpdag{l\sbar}$)~\cite{Tien_PAT_1963, houzet_photon-assisted_2019, diamond_prxq_2022,liu_prl_2024}.
These processes are referred to as photon-assisted QP generation and tunneling (PAT).
For both of the processes described above, the rate of induced qubit state transitions will depend on the specific circuit, Hamiltonian parameters, and the energy of the QPs above the gap, as we will describe in the following sections.

The model presented above is general to any qubit circuit containing JJs.
For concreteness, in this work we consider the fluxonium qubit \cite{manucharyan_fluxonium_2009}, which we model with the Hamiltonian
\begin{equation}
\label{eq: fluxonium_hamiltonian}
\hat{H}_q = 4E_\text{C}(\hat{n}-n_g)^2 - E_\text{J}\cos\hat\varphi + \frac{E_\text{L}}{2}\left(\hat\varphi-\frac{2\pi\Phi_\mathrm{ext}}{\Phi_0}\right)^2,
\end{equation}
where $E_\text{J}$ is the Josephson energy, $E_\text{C}$ is the single-electron charging energy, and $E_\text{L}$ is the inductive energy of the circuit. 
Finally, $\Phi_0=h/2e$ is the superconducting flux quantum.
The small junction of the fluxonium in parallel with the inductor create a superconducting loop, through which an external flux $\Phi_{\text{ext}}$ can be threaded.
We assume the inductor branch of the qubit is realized by a series array of identical junctions (as visualized in Fig.\ref{fig1}(c)) where $E_\text{L} = E_\text{J}^A/N$, with $N$ describing the number of junctions in the array with individual Josephson energy $E_\mathrm{J}^A$.
\begin{figure*}[t!]
    \centering
    \includegraphics[width=\linewidth]{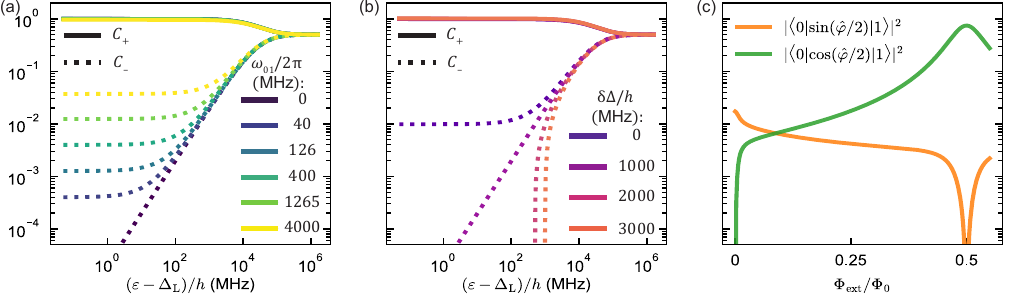}
    \caption{(a) Coherence factors versus QP energy in MHz units for different qubit transition frequencies. All curves assume a symmetric superconducting gap energy ($\delta\Delta = 0$). (b) Coherence factors versus QP energy for different gap asymmetries, both in MHz units, with QP energies measured from $\Delta_\mathrm{L}$. All curves assume a qubit transition frequency $\omega_{01}/2\pi = 1000$ MHz. (c) Squared magnitude of the fluxonium matrix elements for the small JJ QP coupling operators versus applied flux bias.}
    \label{fig2}
\end{figure*}

To numerically model the qubit, relevant for the matrix element calculations we describe later, we solve for the eigenstates of the fluxonium Hamiltonian [Eq.~\ref{eq: fluxonium_hamiltonian}] in the phase basis, discretizing within a range from $-6\pi$ to $6\pi$ with steps $\delta\varphi = 0.04\pi$.
In the following sections, we will model the interactions between a fluxonium and a population of QPs via Fermi's golden rule.

\section{Dissipation from Tunneling of Preexisting Quasiparticles}
Under the assumption that $\hbar\omega_{ij}\ll2\Delta$ and that there are no other sources of pair-breaking radiation in the environment, in this section we consider the qubit transition rates due exclusively to the tunneling of preexisting nonequilibrium QPs. 
Here we will first describe the contribution for QPs tunneling through the fluxonium small JJ.
This allows us to simplify the QP tunneling Hamiltonian in Eq.~\ref{eq: factored qp-qubit interaction term} to
\begin{equation}
\begin{aligned}
    \hqpphi= t\sum_{l,r,s}\bigg[&(u_ru_l-v_rv_l)\cos\frac{\ovphi}{2}\\&+i(u_ru_l+v_rv_l)\sin\frac{\ovphi}{2}\bigg]\qpdag{rs}\qpo{ls}+\mathrm{H.c.}
\end{aligned}
\label{eq: HQP tunneling only}
\end{equation}
%%%%%%%%%%%%%%%%%%%
Defining the BCS coherence factors
\begin{equation}
    C_{\pm} = (u_ru_l\pm v_rv_l)^2 = \frac{1}{2}\left(1 \pm \frac{\Delta_\text{L}\Delta_\text{R}}{\varepsilon_l\varepsilon_r}\right),
\end{equation} 
and applying conservation of energy, $\varepsilon_r = \varepsilon_l + \hbar\omega_{ij}$, we calculate the relative contributions of the terms in Eq.~\ref{eq: HQP tunneling only} for different QP energies [Fig.~\ref{fig2}].
Typically, an assumption is made regarding the energy hierarchy of the system wherein $\varepsilon_{l,r}-\Delta\ll \hbar\omega_{ij}\ll \Delta$, which would allow one approximate $C_-\approx0$ and neglect the cosine term in Eq.~\ref{eq: HQP tunneling only} \cite{catelani_prb_2011, pop_coherent_2014}.

Considering a regime with no gap asymmetry ($\Delta_l=\Delta_r$), we plot the relative magnitude of $C_+$ and $C_-$ versus QP energies for various qubit frequencies $\omega_{01}/2\pi$ [Fig.~\ref{fig2}(a)]. 
We find the prefactors of both terms in Eq.~\ref{eq: HQP tunneling only} are nonzero at all QP energies for nonzero qubit transition frequencies, distinct from the understanding that for QP energies close to $\Delta$, we expect perfect destructive interference of $C_-$.
Simply put, a quasiparticle would not be able to tunnel from gap edge to gap edge while exchanging energy with the qubit.
While $C_-$ may be significantly suppressed relative to $C_+$, the cosine term of Eq.~\ref{eq: HQP tunneling only} can contribute to the overall QP-induced transition rate, which we will describe later.
Inclusion of gap asymmetry across the JJ, however, can suppress $C_-$.
As expected, gap asymmetry larger than $\hbar\omega_{01}$ results in significant suppression of $C_-$ for a wide range of QP energies [Fig.~\ref{fig2}(b)].
Nevertheless, this motivates the inclusion of the full form of Eq.~\ref{eq: HQP tunneling only} when considering the rate of QP tunneling.
This simple observation forms the basis with which we reconsider the intuition around protection of fluxonium qubits against QPs tunneling at the small JJ.

We calculate the limitation that tunneling QPs impose on the energy relaxation time $T_1$ of a fluxonium qubit using Fermi's golden rule.
Throughout this section, we describe numerical computations of QP-energy dependent functions with a discretized QP energy spacing $\delta\varepsilon\approx h\times20$~kHz, which we found sufficient for convergence.
The transition rate is
\begin{equation}
\Gamma_{ij}^{\mathrm{QP},\text{J}}= \frac{2\pi}{\hbar}\sum_{l,r,s}\left|\langle j,r|\hqpphi|i,l\rangle\right|^2\delta_{\epsilon_{i}-\epsilon_{j}+\ve_l-\ve_r}.
\label{eq:FGR}
\end{equation}
%%%%%%%%%%%%%%%%%%%
Here, the ket ~$\ket{i,l}$ represents the initial joint qubit and QP state. 
The sum is over all QP states, with the Kronecker delta enforcing energy conservation.
The energy difference $\epsilon_i - \epsilon_j$ is equal to the qubit transition energy $\hbar\omega_{ij}$.
For simplicity, the formulas below describe QP tunneling from the left to the right side of the JJ (hiding the Hermitian conjugate term of Eq.~\ref{eq: HQP tunneling only}).

\begin{figure*}[t!]
    \centering
    \includegraphics[width=\linewidth]{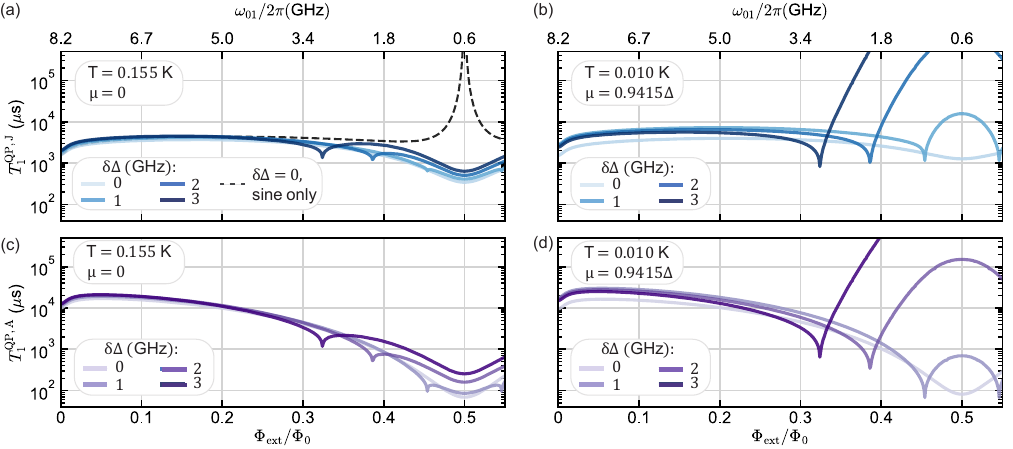}
    \caption{Calculating $T_1$ for QP tunneling through fluxonium junctions for different QP energy distributions, versus applied bias $\Phi_\text{ext}/\Phi_0$. Colors correspond to different gap asymmetries. All curves are calculated with fluxonium Hamiltonian energies $E_{\mathrm{J}}/h = 10.2$~GHz, $E_{\mathrm{C}}/h = 3.6$~GHz, and $E_{\mathrm{L}}/h = 0.46$~GHz. All curves assume $\Delta_\mathrm{L} = 50$ GHz, and $\Delta_\text{R} = \Delta_\text{L} + \delta\Delta$. In the top x-axis, the qubit transition frequencies at the marked flux biases are shown, rounded to the nearest 100 MHz. (a) $T_1^{\mathrm{QP,J}}$ for a ``hot" QP distribution. Black dashed curve corresponds to the $T_1$ prediction including only the $\sin$ term in Eq.~\ref{eq:FGR_tunneling}, for $\delta\Delta = 0$. (b) $T_1^{\mathrm{QP,J}}$ for a ``cold" QP distribution. (c) $T_1^{\mathrm{QP,A}}$ for a ``hot" QP distribution. (d) $T_1^{\mathrm{QP,A}}$ for a ``cold" QP distribution. }
    \label{fig3}
\end{figure*}

Factoring the~$\ovphi$-dependent matrix element out of the sum leaves the following simplified form,

%%%%%%%%%%%%%%%%%%%
\begin{equation}
\begin{aligned}
\label{eq:FGR_tunneling}
\Gamma_{ij}^{\mathrm{QP},\text{J}}=&\left|\langle j|\cos\frac{\ovphi}{2}|i\rangle\right|^2S_-^{\qp,\text{J}}[\omega_{ij}]
\\&+\left|\langle j|\sin\frac{\ovphi}{2}|i\rangle\right|^2S^{\qp,\text{J}}_+[\omega_{ij}].
\end{aligned}
\end{equation}
%%%%%%%%%%%%%%%%%%%
The factors~$S^{\qp,\text{J}}_\pm$ are QP spectral functions. 
They account for the availability and degeneracy of initial and final QP states separated by the qubit energy ~$\varepsilon_r-\varepsilon_l = \hbar\omega_{ij}$, and are defined as
\begin{equation}
\begin{aligned}
S^{\qp,\text{J}}_\pm[\omega_{ij}] =&\frac{2\pi}{\hbar}t^2\int_0^\infty d\ve_l\int_0^\infty d\ve_r  \\
&\times f_l(\ve_l)\left[1-f_r(\ve_r)\right]\\
&\times\nu_l(\ve_l)\nu_r(\ve_r)\nu_n^2\\
&\times (u_ru_l\pm v_rv_l)^2\\
&\times\delta(\epsilon_{i}-\epsilon_{j}+\ve_l-\ve_r).
\end{aligned}
\end{equation}
The first line of the integral accounts for the degeneracy of initial and final states~$l$ and~$r$, where $f_{l,r}(\ve)$ is the QP energy distribution function, and ${\nu_{l,r}(\ve) = \mathrm{Re}\left\{\varepsilon_{l,r}/\sqrt{(\varepsilon_{l,r}^2-\Delta_{l,r}^2)}\right\} }$ is the superconducting density of states normalized by the normal density of states at the Fermi energy $\nu_n$. 
To enforce energy conservation, the Dirac delta function selects only the contribution where $\ve_r=\ve_l+\hbar\omega_{ij}$.
%%%%%%%%%%%%%%%%%%
Following Ref.~\cite{catelani_prb_2011} to recast the tunneling probability into device parameters, accounting for spin degeneracy and the doubling of the density of states in the excitation picture, one arrives at
%%%%%%%%%%%%%%%%%%%
\begin{equation}
\begin{aligned}
\label{eq:gamsqp}
S^{\qp,\text{J}}_\pm[\omega_{ij}] =& \frac{8E_\text{J}}{\pi\hbar\Delta}\int_0^\infty d\ve_l \\
&\times f_l(\ve_l)\left[1-f_r(\ve_l+\hbar\omega_{ij})\right]\\
&\times \nu_l(\ve_l)\nu_r(\ve_l+\hbar\omega_{ij})\\
&\times\left[1\pm\frac{\Delta_l\Delta_r}{\varepsilon_l(\varepsilon_l+\hbar\omega_{ij})} \right].
\end{aligned}
\end{equation}
In this formalism, $T_1^\mathrm{QP,J} =  1/(\Gamma_{01}^{\mathrm{QP},\text{J}} + \Gamma_{10}^{\mathrm{QP},\text{J}})$, where we note the importance of summing over both QP tunneling directions, left to right and right to left.

We also consider QPs tunneling through the JJ-array inductor contributing to the energy relaxation $T_1^{\text{QP,A}}$. 
We assume homogenous QP density between the inductor array and around the small JJ. 
To calculate $T_1^{\text{QP,A}}$, we sum the contribution to the relaxation rate from each JJ in the array, where the phase drop of the qubit is distributed evenly as $(\hat{\varphi} - 2\pi\Phi_\mathrm{ext}/\Phi_0)/N$. In terms of the quantities defined above, the array contributions can be written:
\begin{equation}
\begin{aligned}
\Gamma_{ij}^{\mathrm{QP},\text{A}}=& \left|\langle j|\cos\frac{(\hat{\varphi} - 2\pi\Phi_\mathrm{ext}/\Phi_0)}{2N}|i\rangle\right|^2\frac{N^2E_\mathrm{L}}{E_\mathrm{J}}S_{-}^{\qp,\text{J}}[\omega_{ij}]
\\&+\left|\langle j|\sin\frac{(\hat{\varphi} - 2\pi\Phi_\mathrm{ext}/\Phi_0)}{2N}|i\rangle\right|^2\frac{N^2E_\mathrm{L}}{E_\mathrm{J}}S^{\qp,\text{J}}_{+}[\omega_{ij}],
\end{aligned}
\end{equation}
where we sum the contributions from each array JJ, assuming $N = 101$ for all calculations.
We similarly define $T_1^\mathrm{QP,A} =  1/(\Gamma_{01}^{\mathrm{QP},\text{A}} + \Gamma_{10}^{\mathrm{QP},\text{A}})$, where we again note that one needs to consider tunneling in both directions across the JJs when computing numerical transition rates.

We assume Fermi-Dirac distributions describe the QP energy occupation, $f_{l,r}(\varepsilon_{l,r}) = 1/(e^{(\varepsilon_{l,r} - \mu)/k_\text{B}T} + 1)$, where $\mu$ is the chemical potential and $T$ is the effective temperature of the QP bath \cite{Martinis_EnergyDecay_2009, Goldie_2013}.
We consider two different QP energy distributions, each corresponding to a QP density $x_{\text{QP},l} \approx 1\times10^{-7}$ normalized to the density of Cooper pairs, defined by 
\begin{equation}
    x_{\text{QP},l} = \frac{2}{\Delta}\int_0^\infty d\varepsilon_l f_l(\varepsilon_l)\nu_l(\varepsilon_l).
\end{equation}
As a proxy for the QP energy distribution after a high-energy cosmic ray impact event, we consider a ``hot" distribution with $T = 0.155$ K and $\mu = 0$. 
To model the steady state, when we expect the QPs to be well thermalized to the phonon bath \cite{Connolly_2024}, we consider a ``cold" distribution, with $T = 0.01$~K and chemical potential $\mu = 0.9415\Delta_\mathrm{L}$.
These parameters yield the same effective $x_{\text{QP},l}$ as the ``hot" distribution, allowing for quantitative comparison of predicted $T_1$ for these distinct QP energy distributions.

For each QP energy distribution, we evaluate the predicted $T_1$ numerically.
We additionally vary the gap asymmetry energy across the JJ.
We find $T_1$ sensitivity for both QP energy distributions at all fluxonium biases, for a symmetric superconducting gap profile across the circuit JJs [Fig.~\ref{fig3}]. 
For the ``hot" distribution, we predict $T_1$ sensitivity at all qubit biases, with tunneling through the inductor to largely dominating the limitation to $T_1$ over tunneling through the small junction [Fig.~\ref{fig3}(a, c)]. 
Inclusion of gap asymmetry suppresses this limitation as the qubit transition is tuned below the asymmetry energy, but does not eliminate $T_1$ sensitivity to QP tunneling entirely. 
The model predicts an expected resonant enhancement of tunneling from left to right when $\hbar\omega_{01}=\delta\Delta$, which aligns the divergence in the density of states at the gap on either side of the junction, appearing here as dips in the predicted $T_1$ \cite{diamond_prxq_2022, marchegiani_quasiparticles_2022}. 
Reemphasizing the message of Fig.~\ref{fig2}, both the sine and cosine terms in $\hqpphi$ contribute to the expected $T_1$ at all $\Phi_\mathrm{ext}$ and $\delta\Delta$.

The ``cold" distribution shows a distinct trend for qubit transition energies below 
$\delta\Delta$~[Fig.~\ref{fig3}(b,d)], exhibiting some protection. 
This can be understood from a density of states picture; for a ``cold" QP energy distribution, the QPs predominantly reside in the the lower-gap superconductor, with the high-gap side of the JJ relatively unoccupied. 
The qubit in its excited state can supply QPs with enough energy to tunnel to the high-gap side, so long as $\hbar\omega_{01}\gtrsim\delta\Delta$. 
As $\hbar\omega_{01}$ is tuned below $\delta\Delta$, low-gap-side QPs lack the energy to tunnel to the high-gap side, suppressing tunneling in one direction.

High-gap-side QPs could, in principle, tunnel to the low-gap-side regardless of the magnitude of~$\hbar\omega_{01}$. 
However, the ``cold" QP energy distribution does not appreciably populate the high gap electrode's QP states, resulting in low rates of tunneling in both directions. 
The predicted $T_1$ is commensurately large in this regime, with larger gap asymmetries showing suppressed $T_1$ sensitivity for a wider range of qubit frequencies. 
These predictions are consistent with results in the community showing that gap engineering can result in $T_1$ protection from low energy QP tunneling processes~\cite{diamond_prxq_2022, Connolly_2024}. 
Increasing $\delta\Delta$ relative to the width of the QP energy distribution will further suppress QP-induced errors. 
We also note here that the predicted rate of errors from QP-tunneling within the JJ-array inductor is expected to outweigh tunneling through the small JJ for both energy distributions considered.

%%%%%%%%%%%%%%%%
%%%%%%%%%%%%%%%%%%%
\section{Dissipation from Photon-assisted tunneling}
In the presence of environmental radiation, QP-creation terms in Eq.~\ref{eq: factored qp-qubit interaction term} can become important.
Here we consider $T_1$ due to those processes, accounting for stray electromagnetic radiation in the environment with energy~$\hbar\omega_{p}\gtrsim2\Delta$\cite{houzet_photon-assisted_2019, Tien_PAT_1963}.

\begin{figure}[t!]
    \centering
    \includegraphics[width=\linewidth]{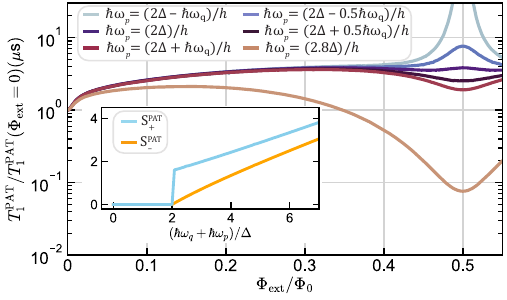}
    \caption{Trend in $T_1$ for PAT processes  at the fluxonium small JJ for different drive frequencies $\omega_p$, relative to the qubit frequency where we define $\hbar\omega_q = \hbar\omega_{01}(\Phi_\text{ext}/\Phi_0 = 0.5)$. The $T_1$ for each curve is plotted normalized by its value at $\Phi_\mathrm{ext}/\Phi_0 = 0$. The inset shows the contributions of the spectral functions $S_\pm^\text{PAT}$ versus the total PAT process energy ($\hbar\omega_p + \hbar\omega_q$), normalized the to superconducting gap energy.}
    \label{fig4}
\end{figure}

To model the effect of an ac electric field coupled to the qubit-QP system, we follow Ref.~\cite{houzet_photon-assisted_2019} by adding a term $\hbar\omega_p\hat{b}_p^\dagger\hat{b}_p$ to our system Hamiltonian, where $\hat{b}_p^\dagger$ and $\hat{b}_p$ are creation and annihilation operators of a pair-breaking photon. 
We additionally make the substitution $\hat{\varphi}=\hat{\varphi}+\phi_p(\hat{b}_p + \hat{b}_p^\dagger)$, where $\phi_p$ is the amplitude of the zero-point fluctuations in phase of the qubit induced by the photon.
For weak coupling, we assume $\phi_p\ll 1$.

With this transformation, the QP-qubit interaction Hamiltonian for PAT processes becomes
%%%%%%%%%%%%%%%%
\begin{equation}
\begin{aligned}
    \hat{H}_\mathrm{PAT} = &t\frac{i\phi_p}{2}(\hat{b}_p + \hat{b}_p^\dagger)\\
    &\times\sum_{l,r,s}\Bigg\{\bigg[(u_ru_l-v_rv_l)\sin\frac{\ovphi}{2}\\
    &\ \ \ \ \ \ \ \ \ \ \ \ \ +i(u_ru_l+v_rv_l)\cos\frac{\ovphi}{2}\bigg]\qpdag{rs}\qpo{ls}\\
    &\ \ \ \ \ \ \ \ \ \ \ +\bigg[(u_rv_l-v_ru_l)\cos\frac{\ovphi}{2}\\
    &\ \ \ \ \ \ \ \ \ \ \ \ \ \ \ \ +i(u_rv_l+v_ru_l)\sin\frac{\ovphi}{2}\bigg]\qpdag{rs}\qpdag{l\sbar}\Bigg\},\\
\end{aligned}
\end{equation}
neglecting the Hermitian conjugate term.
The first term will be suppressed relative to the pair-tunneling by a factor~$\sim1/\xqp$, so we will neglect it.
The PAT process produces two QPs, one on each side of the junction, with opposite spin~$\qpdag{rs}$ and~$\qpdag{l\sbar}$. 

Here again we use Fermi's golden rule to calculate the transition rate due to PAT. 
The relaxation rate can be written as
%%%%%%%%%%%%%%%%%%%
\begin{equation}
\begin{aligned}
\Gamma_{ij}^\mathrm{PAT}=&\left|\langle j|\cos\frac{\ovphi}{2}|i\rangle\right|^2S_-^\mathrm{PAT}[\omega_{ij}]
\\&+\left|\langle j|\sin\frac{\ovphi}{2}|i\rangle\right|^2S^\mathrm{PAT}_+[\omega_{ij}]
\end{aligned}
\end{equation}
%%%%%%%%%%%%%%%%%%%
with the PAT spectral functions:
\begin{equation}
\begin{aligned}
S^\pat_\pm[\omega_{ij}] =& \frac{2\pi}{\hbar}\left(t\frac{\phi_p}{2}\right)^2\int_0^\infty d\ve_l\int_0^\infty d\ve_r \\
&\times\left[1- f_l(\ve_l)\right]\left[1- f_r(\ve_r)\right]\\
&\times \nu_l(\ve_l)\nu_r(\ve_r)\nu_n^2\\
&\times (u_rv_l\pm v_ru_l)^2\\
&\times\delta(\hbar\omega_p-\ve_l-\ve_r+\epsilon_{i}-\epsilon_{j}).
\end{aligned}
\end{equation}
%%%%%%%%%%%%%%%%%%
Assuming that there are very few preexisting QPs, we approximate~$\left[1-\nlb f(\ve_{l,r})\right]\approx\nlb1$. 
Energy conservation is ensured by setting $\ve_r=\hbar\omega_p+\hbar\omega_{ij}-\ve_l$. 
All together, the PAT spectral functions are
\begin{equation}
\begin{aligned}
S^\pat_\pm[\omega_{ij}] =& \frac{2E_\text{J}\phi_p^2}{\pi\hbar\Delta}\int_0^\infty d\ve_l \\
&\times\nu_l(\ve_l)\nu_r(\hbar\omega_p+\hbar\omega_{ij}-\ve_l)\\
&\times \left[1\pm\frac{\Delta_l\Delta_r}{\varepsilon_l(\hbar\omega_p+\hbar\omega_{ij}-\ve_l)} \right].
\end{aligned}
\label{eq:PATGammaS}
\end{equation}

We calculate $T_1$ for photons coupling to the fluxonium small JJ, where we assume a symmetric gap energy across the JJ (or equivalently, a gap asymmetry energy $\ll \hbar\omega_p$).
This is motivated by typical realizations of the qubit, in which large capacitor pads connect to this small JJ, increasing the absorption cross section for pair breaking radiation~\cite{liu_prl_2024}.
Additionally, we assume for simplicity that the pair-breaking radiation is monochromatic, motivated by the results of \cite{houzet_photon-assisted_2019,liu_prl_2024}, which suggest that there are regimes in which this may be approximately true.
We normalize the calculated the $T_1$ to its value at $\Phi_\text{ext} = 0$ to avoid assumptions in calculating $\phi_p$, which will depend on specific geometric details of the electromagnetic environment, and could in principle introduce additional dependence on $\Phi_\mathrm{ext}$.
We investigate the trend in the PAT-induced $T_1$ for different monochromatic environmental drive frequencies referenced to the qubit transition frequency at $\Phi_\mathrm{ext} = 0.5$, $\omega_q/2\pi$. 
For numerical results presented in this section, we adaptively modify $\delta\varepsilon$ to smooth out discontinuities introduced by the gap edge.

The predicted PAT-induced $T_1$ as a function of  pair-breaking frequency is shown in Fig~\ref{fig4}. 
When the total energy transferred in the PAT process ($\hbar\omega_p + \hbar\omega_q$) approaches $2\Delta$ from above, we observe a relative increase in $T_1$ due to this process. 
In this limit, the terms associated with $S_-^\text{PAT}$ are very nearly zero, and the fluxonium qubit is protected via the symmetry properties of the $\sin\hat{\varphi}/2$ matrix element which suppresses the contribution from $S_+^\text{PAT}$. 
Accordingly, we observe decreased $T_1$ sensitivity to this mechanism specifically when considering pair breaking radiation in an extremely small window around $2\Delta$ in energy.
However, for photons with energy even slightly above $2\Delta$, we observe $T_1$ to be sensitive to PAT processes at all qubit biases.
Though we only consider monochromatic radiation here, from these calculations we hypothesize that fluxonium qubits would not be protected against PAT processes for realistic environmental conditions.

When $\hbar\omega_p< 2\Delta + \hbar\omega_q$, the PAT process will exclusively cause relaxation errors in the qubit. 
Above this energy, the PAT process can additionally excite the qubit. 
For PAT frequencies inducing excitation and relaxation in the qubit, we predict a trend in $T_1$ similar to the that corresponding to preexisting QPs tunneling through the inductor junctions with $\delta\Delta = 0$ in Fig.~\ref{fig3}.

\section{Conclusions}
QPs created by high-energy radiation events motivate revisiting the often-used assumption that the QP-qubit system can be described by the energy hierarchy $\varepsilon-\Delta\ll \hbar\omega_{ij}\ll \Delta$.
Lifting this assumption, we numerically simulate the QP-induced $T_1$ for a fluxonium qubit.
In considering the $T_1$ due to a nonequilibrium population of QPs regardless of whether they are described by either a ``hot" or ``cold" energy distribution, we generally find the fluxonium to exhibit $T_1$ sensitivity to QPs, but in ways that deviate from commonly held intuition.

For ``hot," high-energy QP distributions, this model predicts $T_1$ sensitivity at all fluxonium biases.
The model predicts insensitivity to ``cold," low energy QPs, but only when considering JJs with asymmetric superconducting gaps, when the qubit transition energy is below $\delta\Delta$. 
These findings contradict intuition that the fluxonium and other qubit circuits are intrinsically insensitive to QP-induced errors at their operating points, and reiterate that engineering a favorable superconducting gap profile decreases $T_1$ sensitivity to QPs.
Furthermore, complex gap engineering strategies between the qubit circuit and surrounding ground plane can reduce QP occupation in the JJ electrodes, further suppressing tunneling~\cite{Kurilovich_2026,pinckney2026characterization}.

We further apply an existing theoretical formalism~\cite{houzet_photon-assisted_2019} to consider the PAT processes in fluxonium and investigate the $T_1$ for different monochromatic pair-breaking photon frequencies.
We predict protection at $\Phi_\text{ext}/\Phi_0~=~0.5$ only for processes with total energy transfer of approximately $2\Delta$.
Above this energy scale, we predict fluxonium sensitivity to PAT processes at all biases, with sensitivity increasing with the total energy of the process, indicating PAT processes are just as much a necessary consideration for the fluxonium as they are for transmon qubits~\cite{liu_prl_2024} and could contribute to elevated excited state populations even if the rate of absorption is low~\cite{serniak_hot_2018, diamond_prxq_2022}.

The implications of these findings extend to qubits beyond the fluxonium considered here.
These findings are also relevant for any qubit circuits for which intrinsic suppression of QP-induced dissipation relies on the picture of destructive interference of electron-like and hole-like QP tunneling processes (and associated arguments of wavefunction symmetry~\cite{pop_coherent_2014}) at specific phase biases~(for example,~Refs.~\cite{Orlando_1999,manucharyan_fluxonium_2009,smith2020superconducting} among others).
This motivates the pursuit of circuits with very nearly degenerate ground states or disjoint support to further suppress QP-induced depolarization errors~\cite{Brooks_zeropi_2013,kalashnikov_bifluxon_2020, hays_harmonium_2025}, which gives rise to additional experimental challenges.  

\textit{Note added.}
We recently became aware of related, experimental work on QP-induced dissipation in fluxonium qubits \cite{litskevich2026qpfluxonium}.

\section{Acknowledgments}
The authors thank M. Houzet for many valued technical exchanges, R. DePencier Piñero, J. M. Gertler, and M. T. Randeria for discussions and contributions to the simulation infrastructure, and L. I. Glazman, J. A. Grover, A. J. Kerman, V. D. Kurilovich, W. D. Oliver, I. V. Pechenezhskiy, and K. L. Tiwari for technical feedback and discussions. 
This research was funded under Air Force Contract No. FA8702-15-D-0001.
The views and conclusions contained herein are those of the authors and should not be interpreted as necessarily representing the official policies or endorsements, either expressed or implied, of the U.S. Air Force or the U.S. Government.

\bibliography{bibliography.bib}
\end{document}